\documentclass[twocolumn,reprint,amsmath,amssymb, aps,prl]{revtex4-2}
\usepackage[T1]{fontenc}
\usepackage[latin9]{inputenc}
\usepackage{color}
\usepackage{float}

\makeatletter

\usepackage{amsfonts}
\usepackage{latexsym}
\usepackage{epsfig}
\usepackage{graphicx}
\usepackage{tikz}
\definecolor{greatblue}{RGB}{40,120,181}
\definecolor{greatred}{RGB}{200,36,35}
\usepackage[colorlinks,linkcolor=greatblue,anchorcolor=blue,citecolor=greatred]{hyperref}

\usepackage{dcolumn}
\usepackage{bm}

\makeatother

\usepackage[english]{babel}
\begin{document}
\preprint{preprintnumbers}{CTP-SCU/2025004}
\title{Timelike Entanglement Entropy Revisited}

\author{Xin Jiang}  
\email{domoki@stu.scu.edu.cn}
\affiliation{College of Physics, Sichuan University, Chengdu, 610065, China}



\author{Haitang Yang}
\email{hyanga@scu.edu.cn}
\affiliation{College of Physics, Sichuan University, Chengdu, 610065, China}


\begin{abstract}
We present an operator-algebraic  definition for timelike entanglement entropy in QFT under a few mild postulates. This rigorously defined timelike entanglement entropy is real-valued due to the timelike tube theorem. We further demonstrate why the  timelike entanglement entropy  should be real-valued from both path integral argument and holography perspective.
\end{abstract}

\maketitle

\section{Introduction}

As an intrinsic ingredient in quantum theory, entanglement entropy 
is increasingly playing a central role in 
quantum field theory (QFT) and quantum gravity. 
The standard definition of entanglement entropy in a pure state is the von Neumann 
entropy $S_{\text{vN}}=-\mathrm{Tr}\rho_A\log\rho_A$
for a reduced density matrix $\rho_A = \mathrm{Tr}_{A^c}\rho$ which is obtained after tracing out the 
complement of  $A$. Usually, the entanglement entropy in QFT  is defined for a spacelike region. 
Recently,  entanglement entropy for
``timelike regions'' has been introduced from different theoretical motivations
\citep{Wang:2018jva,Liu:2022ugc,Doi:2022iyj,Doi:2023zaf,Milekhin:2025ycm}.
Two  seemingly conceptually distinct configurations of 
timelike entanglement are proposed. The configuration A proposed in 
\citep{Wang:2018jva,Liu:2022ugc,Doi:2022iyj,Doi:2023zaf}
addresses the entanglement between an interval  and the remaining  segment on the 
temporal axis, as shown in Figure \ref{fig:timelike-envelope}.  
The configuration B proposed in
\citep{Milekhin:2025ycm} investigates the entanglement between two timelike-separated spacelike regions.
We focus on configuration A in the present work, 
and reserve the discussion of configuration B for future studies.
An intrinsic definition of timelike entanglement entropy, however,
is still unclear. In particular, two key questions remain unanswered:
What exactly is the timelike subsystem in QFT? And how to define entanglement
entropy of the timelike subsystem? 

A local field operator $\phi(x)$ in QFT cannot be considered a true
observable, as $\phi(x)$ takes states out of the Hilbert space $\mathcal{H}$
when acting on them. This is manifested by the singularity appears 
in the operator product expansion (OPE) of $\phi(x)\phi(y)$.
To obtain a well-defined Hilbert space operator,
one might smear $\phi(x)$ over an open region $\mathcal{O}$, yielding
$\Phi_{f}=\int_{\mathcal{O}}\mathrm{d}\mu\,f(x)\phi(x)$, where $f$
is a smooth test function. However, smearing field operators over
a spatial region $V$ does not always produce a true operator. For
instance, in QCD, smearing in space fails
due to the non-integrable singularities in the OPE
\citep{Witten:2023qsv}. Instead, Borchers proved \citep{Borchers1964}
that it suffices to smear out the field operator in \emph{real time},
\begin{equation}
\Phi_{f}(\vec{x}_{0})=\int_{\mathcal{T}}\mathrm{d}t\,f(t)\phi(x),
\end{equation}
where the test-function $f(t)$ is supported on a timelike interval,
\begin{equation}
\mathcal{T}=\left\{ x=(t,\vec{x});\vert t\vert<\frac{T}{2},\vec{x}=\vec{x}_{0}\right\} .
\end{equation}

Smearing in real time always converts $\phi(t,\vec{x})$ into true
operators, enabling the definition of an algebra $\mathcal{A}$ generated
by bounded functions of operators supported on $\mathcal{T}$. In
the framework of algebraic QFT \citep{Haag1964,Haag:1996hvx}, each
open region in Minkowski spacetime is associated with an algebra $\mathcal{A}$
of observables, constituting a subsystem. We can thus view the algebra
$\mathcal{A}(\mathcal{T})$, supported on the timelike interval $\mathcal{T}$,
as a proper timelike subsystem in QFT across general dimensions. 
In this paper, the class of algebras we consider is sometimes referred to in the literature as the \textit{additive algebra} \citep{Casini:2020rgj, Casini:2021zgr}. By this, we mean that the theory contains no extended operators, such as nontrivial Wilson loops, that are not generated by local fields. Under this assumption, the algebra $\mathcal{A}(\mathcal{O})$ exhausts all local observables measurable within the region $\mathcal{O}$. In this sense, an algebraic timelike subsystem $\mathcal{A}(\mathcal{T})$ is physically meaningful, since it is generated by observables localized along an observer's worldline $\mathcal{T}$.

Defining the density matrix and von Neumann entropy in QFT is not
straightforward. Unlike the Type $\mathrm{I}$ von Neumann algebra
in  quantum mechanics, the algebra $\mathcal{A}(\mathcal{O})$
associated with a causal diamond (or double cone) $\mathcal{O}$ in
QFT is typically not isomorphic to $\mathfrak{B}(\mathcal{H}_{\mathcal{O}})$,
the algebra of all bounded linear operators on some Hilbert space.
For a causal diamond $\mathcal{O}$ and its causal complement $\mathcal{O}^{\prime}$,
the factorization of the vacuum Hilbert space $\mathcal{H}=\mathcal{H}_{\mathcal{O}}\otimes\mathcal{H}_{\mathcal{O^{\prime}}}$
does not exist. Notwithstanding these challenges, for two centric causal diamonds
$\mathcal{O}$ and $\widetilde{\mathcal{O}}$, the \textit{split property}, originally conjectured by Borchers \citep{Haag:1996hvx}  and studied in detail by Doplicher and Longo \citep{Doplicher:1984zz}, guarantees the existence of an intermediate
Type $\mathrm{I}$ factor $\mathcal{N}$ such that
\begin{equation}
\mathcal{A}(\mathcal{O})\subset\mathcal{N}\subset\mathcal{A}(\widetilde{\mathcal{O}}),\quad\mathcal{O}\subset\widetilde{\mathcal{O}}.
\end{equation}
The vacuum Hilbert space $\mathcal{H}$ can then be factorized as $\mathcal{H}=\mathcal{H}_{\mathcal{N}}\otimes\mathcal{H}_{\mathcal{N^{\prime}}}$,
where $\mathcal{N}^{\prime}$ is the commutant of $\mathcal{N}$,
defined as the set of all bounded operators on $\mathcal{H}$ that
commute with $\mathcal{N}$. It has been suggested \citep{Longo:2019pjj}
that the vacuum von Neumann entropy of $\mathcal{A}$ with respect
to $\mathcal{O}$ and $\widetilde{\mathcal{O}}$ could be defined
as
\begin{equation}
S_{\mathcal{A}}(\mathcal{O},\widetilde{\mathcal{O}})=-\mathrm{Tr}\rho_{\mathcal{N}}\log\rho_{\mathcal{N}},
\end{equation}
where $\mathcal{N}$ is chosen as a canonical intermediate Type $\mathrm{I}$
factor \citep{Doplicher:1984zz}, and $\rho{}_{\mathcal{N}}$ is the
vacuum density matrix corresponding to $\mathcal{N}$. 
We emphasize that the definition of entropic quantities in QFT via the split property is far from unique. A number of distinct approaches have appeared in the literature; see \citep{Narnhofer:2002Entanglement, Casini:2009Reduced, Otani:2017Towards, Hollands:2018Entanglement, Witten:2018zxz, Dutta:2019gen, Kudler-Flam:2023hkl}.
In our analysis,
it is sufficient to consider the limit,
\begin{equation}
S(\mathcal{O}):=\lim_{\widetilde{\mathcal{O}}\rightarrow\mathcal{\mathcal{O}}}S_{\mathcal{A}}(\mathcal{O},\widetilde{\mathcal{O}}).\label{eq:vN-entropy}
\end{equation}
Here, $S(\mathcal{O})$ would capture the universal UV divergence
when $\widetilde{\mathcal{O}}$ approaches to $\mathcal{\mathcal{O}}$.
Therefore, in QFT, $S(\mathcal{O})$ can be interpreted as the entanglement
entropy for the algebra $\mathcal{A}(\mathcal{O})$ of operators in
a causal diamond $\mathcal{O}$.
The split property is often taken to characterize physically acceptable theories, but it is not guaranteed in a general quantum field theory. In this paper, we shall always assume that the QFTs of interest satisfy the split property. It should be noted that the split property can be derived from the Buchholz--Wichmann nuclearity condition, which constrains the admissible number of local degrees of freedom \citep{Buchholz:1986Causal, Buchholz:1987Universal}. This notion of nuclearity has also been formulated in terms of the modular operator of local algebras \citep{Buchholz:1990NuclearI, Buchholz:1990NuclearII}. For further discussions of the split property and nuclearity in conformal field theory, see \citep{Buchholz:2007Nuclearity, Morinelli:2018Conformal, Dutta:2019gen, Ceyhan:2025qrj}; for the corresponding developments in quantum field theory on curved spacetime, see \citep{Verch:1993fs, DAntoni:2006Dirac, Fewster:2015Split, Cao:2025els}.

\section{Timelike tube theorem and timelike entanglement entropy}

\begin{figure}[h]
\includegraphics[width=0.25\textwidth]{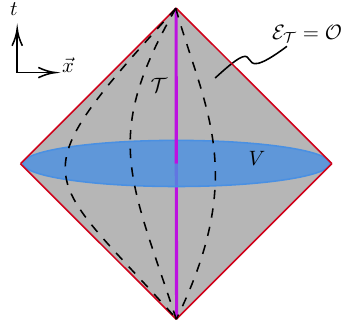}
\caption{The timelike envelope $\mathcal{E}_{\mathcal{T}}$ (the gray shaded
region) consists of all points that can be reached by deforming the
timelike interval $\mathcal{T}$ (the purple line) to a family of
timelike curves (black dashed lines), which is equivalent to a causal
diamond $\mathcal{O}$. All points at $t=0$ constitutes a spacelike
ball $V$ (the blue region).\label{fig:timelike-envelope}}
\end{figure}

To define the entanglement entropy for the algebra $\mathcal{A}(\mathcal{T})$
of operators on a timelike interval $\mathcal{T}$, we first
investigate the relationship between the algebra $\ensuremath{\mathcal{A}(\mathcal{T})}$
and the algebra $\mathcal{A}(\mathcal{O})$ associated with a causal
diamond $\mathcal{O}$. In Minkowski spacetime, this question has been addressed
by the timelike tube theorem \citep{Borchers1961,Araki1963}. Starting
with the timelike interval $\mathcal{T}$, its timelike envelope $\mathcal{E}_{\mathcal{T}}$
consists of all points reachable by deforming $\mathcal{T}$ along
a family of timelike curves while keeping it fixed near its endpoints.
In the following discussion, we restrict our analysis to favorable
cases where the timelike envelope $\mathcal{E}_{\mathcal{T}}$ coincides
with a causal diamond $\mathcal{O}$, which is defined as the intersection of the past
and future of $\mathcal{T}$,
\begin{equation}
\mathcal{E}_{\mathcal{T}}=\mathcal{O}=J^{+}(\mathcal{T})\cap J^{-}(\mathcal{T}),
\end{equation}
as illustrated in Figure \ref{fig:timelike-envelope}.

The timelike tube theorem asserts that the algebra of operators on
a timelike interval $\mathcal{T}$ is identical to the algebra of
operators in its timelike envelope $\mathcal{E}_{\mathcal{T}}$, i.e.,
\begin{equation}
\mathcal{A}(\mathcal{T}) = \mathcal{A}(\mathcal{E}_{\mathcal{T}}).\label{eq:time-tube}
\end{equation}
This relation indicates that the timelike entanglement entropy can be
defined as
\begin{equation}
S(\mathcal{T}):=S(\mathcal{E}_{\mathcal{T}}),\label{eq:timeEE}
\end{equation}
where $S(\mathcal{E}_{\mathcal{T}})$ is the von Neumann entropy defined
in (\ref{eq:vN-entropy}). This definition of timelike entanglement
entropy relies on the principle that identical algebras should yield
identical entanglement entropies. We emphasize that the timelike entanglement
entropy depends only on the timelike interval $\mathcal{T}$, since
the timelike envelope $\mathcal{E}_{\mathcal{T}}$ is uniquely determined
by $\mathcal{T}$. Notice that $\mathcal{E}_{\mathcal{T}}$ is the
domain of dependence of a spacelike ball $V$ at $t=0$, see Figure
\ref{fig:timelike-envelope}. 

Observable algebra in the QFT of our interest should be equipped with the so-called primitive causality \citep{Haag:1962}, which assumes a dynamical equation, or in algebraic terms a time-slice property, such that the observables on a Cauchy slice determine those in the spacetime region causally dependent on that slice. 
The reader could find more helpful discussions on the local algebras in the nice review \citep{Witten:2018zxz}.
Consequently, the algebra $\mathcal{A}(\mathcal{E}_{\mathcal{T}})$
is equivalent to the algebra $\mathcal{A}(V)$ of operators in an
arbitrarily small spacetime neighborhood of $V$, up to unitary time
evolution. Several known results for $S(V)$,
or equivalently $S(\mathcal{E}_{\mathcal{T}})$, have been derived using the Euclidean path
integral approach \citep{Callan:1994py} and holography \citep{Ryu:2006bv,Ryu:2006ef}.

It is ready to calculate the timelike entanglement entropy in CFT 
for a timelike interval,
\begin{equation}
\mathcal{T}=\left\{ x=(t,\vec{x});\vert t\vert<\frac{T}{2},\vec{x}=\vec{x}_{0}\right\} .
\end{equation}
In particular for a $(1+1)$-dimensional zero-temperature CFT on the
spacetime $\mathbb{R}^{2}$, the spacelike ball $V$ reduces to a
spacelike interval of the length $T$. The entanglement entropy
of $V$ is known as \citep{Calabrese:2004eu,Calabrese:2009qy}:
\begin{equation}
S(V)=\frac{c}{3}\log\frac{T}{\epsilon},
\end{equation}
where $c$ is the central charge and $\epsilon$ is the UV cutoff.
Following the definition (\ref{eq:timeEE}), the timelike entanglement
entropy is then given by
\begin{equation}
S(\mathcal{T})=S(V)=\frac{c}{3}\log\frac{T}{\epsilon},
\end{equation}
in agreement with earlier results in \citep{Liu:2022ugc}. This
result can be straightforwardly extended to finite temperatures:
\begin{equation}
S(\mathcal{T})=\frac{c}{3}\log\left(\frac{\beta}{\pi\epsilon}\sinh\frac{\pi T}{\beta}\right),
\end{equation}
where $\beta$ is the inverse temperature. For a finite-size CFT$_2$ on
$\mathbb{R}\times S^{1}$, the timelike entanglement entropy becomes
\begin{equation}
S(\mathcal{T})=\frac{c}{3}\log\left(\frac{L}{\pi\epsilon}\sin\frac{\pi T}{L}\right),
\end{equation}
with restriction $T<L$, where $L$ is the total length of the
spatial circle $S^{1}$. However, caution is required when $T\ge L$,
as the timelike envelope $\mathcal{E}_{\mathcal{T}}$ may not always
correspond to a causal diamond. More generally, for QFT$_{d+1}$,
$V$ is a $d$-dimensional spacelike ball with \ radius $R=T/2$,
we still have
\begin{equation}
S(\mathcal{T})=S(V),\label{eq:timeEE-spaceEE}
\end{equation}
since the timelike tube theorem (\ref{eq:time-tube}) holds in general
dimensions.

\section{Path integral and holography}

Recent literature has predominantly considered a complex-valued timelike
entanglement entropy,  whereas our proposal
yields a real-valued result. It is therefore important to resolve
this contradiction. We  work this out from both path integral argument and 
holography perspective.
We  stress again that 
we are going to focus on configuration A as depicted in  Figure \ref{fig:timelike-envelope}.

\subsection{Path integral argument}

\begin{figure}[h]
	\includegraphics[width=0.5\textwidth]{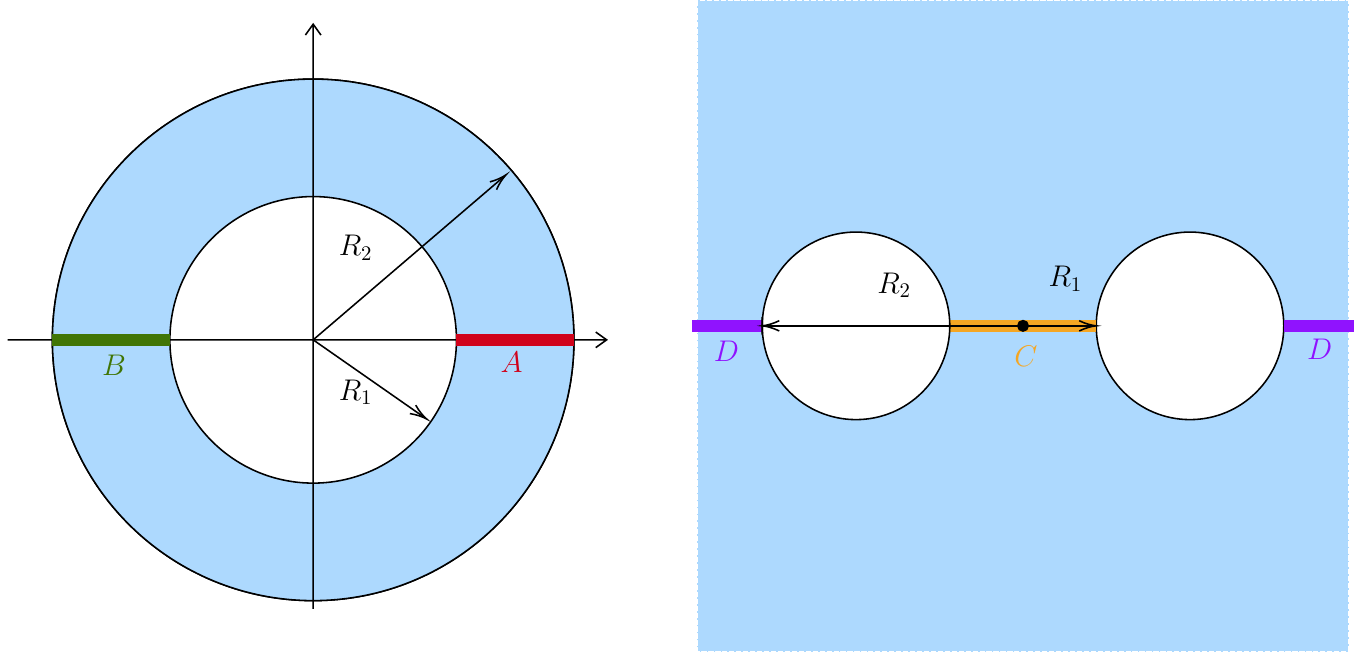}
	\caption{The density matrix for disjoint symmetric segments $A(C)$ and $B(D)$, both of which are reflection-symmetric. The usual 
		adjacent configurations with divergent entanglement entropies $c/3 \log \ell/\epsilon$ or  
		$c/6 \log \xi/\epsilon$
		are obtained by taking simple limits. 
		\label{fig:annuls}}
\end{figure}
Let us consider the density matrix $\rho_{AB}$ in a two-dimensional CFT on an annulus, first studied in detail by  Cardy \citep{Cardy1986}, where $A$ and $B$ are two disjoint cuts in the path integral, as shown in Figure \ref{fig:annuls}. 
Strictly speaking, the von Neumann entropy of entanglement between $A$ and $B$ should contain a non-universal contribution arising from operator insertions, which is encoded in the annulus partition function \citep{Cardy:2004hm}. However, this partition function is dominated by the vacuum sector in the two limits of interest to us: the adjacent limit, in which $A$ and $B$ become neighboring intervals \citep{Cardy:2016fqc}, and the large-central-charge limit associated  to the holography \citep{Brown1986}. In both cases, the vacuum contribution provides the leading behavior. To this end, the vacuum entanglement entropy, namely the contribution in the absence of operator insertions, between two disjoint symmetric spacelike intervals $A$ and $B$ can be written as \citep{Jiang:2024ijx, Jiang:2024hjz, Jiang:2025tqu, Jiang:2025jnk}
\begin{equation}
S_{\rm vN}(A:B)= \frac{c}{6}\log\frac{R_2}{R_1}.
\label{eq:Mixed_EE}
\end{equation}
The entanglement entropy between   $C$ and $D$ are also  given in \citep{Jiang:2024ijx,Jiang:2025tqu}
after applying conformal transformations,
\begin{equation}
S_{\rm vN}(C:D)= \frac{c}{3}\log\frac{\sqrt R_2 + \sqrt R_1}{\sqrt R_2 - \sqrt R_1}.
\label{eq:S_CD}
\end{equation}
We emphasize that the quantity considered here is not the entanglement entropy of a union of disconnected regions \(AB\), which may be  viewed as a subsystem of a larger state such as \(\rho_{ABCD}\). Instead, after the subtraction procedure \citep{Jiang:2024ijx}, one obtains a pure state \(\rho_{AB}\)  in the annulus CFT. The entropic quantity we study is therefore the von Neumann entropy of the subsystem \(A\) in the pure state \(\rho_{AB}\). In the vacuum-dominated limits relevant for our discussion, this quantity is governed by the annulus partition function and gives rise to the expressions above. Our construction is thus different from the standard problem of the entanglement entropy of two generic disjoint intervals in a CFT vacuum state. For this reason, the familiar non-universal dependence of the R\'{e}nyi entropies for generic disjoint intervals on the full operator content does not apply here in the same direct way \cite{Calabrese:2009ez,Calabrese:2010he}.
The usual divergent entanglement entropies between adjacent regions are obtained 
by taking simple adjacent limits of this disjoint system as follows, 
\begin{itemize}
\item Under the limit $R_2-R_1 =\epsilon\to 0$
\begin{equation}
S_{\rm vN}(C:D) = \frac{c}{3}\log \frac{\ell}{\epsilon} + \hbox{finite terms},
\label{eq:S_CD_adj}
\end{equation}
where $\ell = 2R_1$ is the  length of the entangling interval $C$.
\item Under the limit $R_1=\epsilon\to 0$ and $R_2=\xi \to \infty$,
\begin{equation}
S_{\rm vN}(A:B) = \frac{c}{6}\log \frac{\xi}{\epsilon} + \hbox{finite terms},
\label{eq:S_AB_adj}
\end{equation}
which represents the  half-space entanglement entropy.
\end{itemize}
It is evident that the path integrals on this single annulus $Z_1$ and replica $n-$folded 
annulus $Z_n$ are both independent of the locations of $A$ and $B$ on the annulus.
 Consequently, the entanglement entropy between any two disjoint segments is always given 
by either equation (\ref{eq:Mixed_EE}) or (\ref{eq:S_CD}).
Asymmetric configurations can be generated via conformal transformations, 
which do not introduce complex numbers.

We now consider the case that  $A$ and $B$ are both timelike. This is achieved by 
performing a Wick rotation on both  $R_2$ and $R_1$. We of course still get the real-valued 
entanglement entropies given in equations (\ref{eq:Mixed_EE}) and (\ref{eq:S_CD}). 
For the adjacent limits, it is clear that
Wick rotations must be applied simultaneously to both
$\ell$ and $\epsilon$ in  equation (\ref{eq:S_CD_adj}); 
and  to both $\xi$ and $\epsilon$
in  equation (\ref{eq:S_AB_adj}), which surely leads to real-valued results.

However, if we  Wick rotate only $\ell$ in equation (\ref{eq:S_CD_adj}),
or only $\xi$ in  equation (\ref{eq:S_AB_adj}),
while keeping  $\epsilon$ fixed in both scenarios, spurious complex numbers emerge.
This is precisely the approach adopted in previous works, 
where the UV cutoff \(\epsilon\) is introduced ad hoc and arbitrarily set to a real value.


%

\subsection{Holography perspective}


Without loss of generality, consider the entanglement entropy of a single
timelike interval $\mathcal{T}_{1}=\left\{ 0<t<1,x=0\right\} $ in
zero-temperature CFT$_2$. This interval can be
mapped onto the temporal half-axis $\mathcal{T}_{\text{h}}=\left\{ t>0,x=0\right\} $
via a special conformal transformation,
\begin{equation}
x^{\mu}\rightarrow\frac{x^{\mu}-b^{\mu}x^{2}}{1-2(b\cdot x)+b^{2}x^{2}},
\end{equation}
with a timelike vector  $b^{\mu}=(1,0)$. 
Therefore, the entanglement entropy of any
finite timelike interval $\mathcal{T}$ equals that of the temporal
half-axis $\mathcal{T}_{\text{h}}$ due to conformal symmetry:
\begin{equation}
S(\mathcal{T})=S(\mathcal{T}_{\text{h}}).\label{eq:half-finite-equiv}
\end{equation}
\begin{figure}
\includegraphics{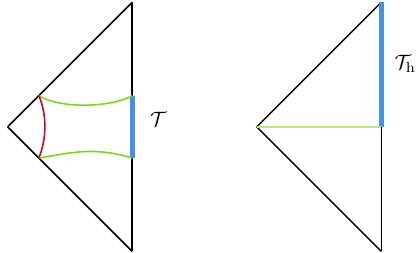}
\caption{Penrose diagrams of Poincar\'{e} patch of AdS$_{3}$ spacetime and
holographic duals of $S(\mathcal{T})$, $S(\mathcal{T}_{\text{h}})$.
In the left panel, the blue line represents a finite timelike interval
$\mathcal{T}$ on the conformal boundary of AdS$_{3}$, the red line
denotes a timelike geodesic, and the two green lines denote spacelike
geodesics. In the right panel, the blue line represents the temporal
half-axis $\mathcal{T}_{\text{h}}$, and the green line denotes a
spacelike geodesic. \label{fig:Penrose-diagrams}}
\end{figure}

It has been conjectured that the holographic dual of $S(\mathcal{T})$
consists of two spacelike geodesics and one timelike geodesic connecting
them in AdS$_{3}$ spacetime, with the metric
\begin{equation}
\mathrm{d}s^{2}=\frac{\mathrm{d}z^{2}+\mathrm{d}x^{\mu}\mathrm{d}x_{\mu}}{z^{2}},
\end{equation}
as illustrated in Figure \ref{fig:Penrose-diagrams}. The timelike
geodesic is argued to contribute an imaginary part to $S(\mathcal{T})$. 

However, this conjecture fails for $S(\mathcal{T}_{\text{h}})$, whose
holographic dual is a single spacelike geodesic from $(z,x^{\mu})=(0,0)$
to $(z,x^{\mu})=(+\infty,0)$, containing no timelike geodesics. 
Thus,
from the holographic perspective, $S(\mathcal{T}_{\text{h}})$ is
purely real-valued, and by the equivalence in (\ref{eq:half-finite-equiv}),
$S(\mathcal{T})$ must also be real-valued.
Moreover, one can easily see that the real-valued  $S(\mathcal{T}_{\text{h}})$  represents
the entanglement entropy between $t\in (0,\infty)$ and $t\in (-\infty,0)$.

\section{Entanglement across time}

It is natural to ask whether the timelike entanglement entropy represents
entanglement across time. The answer appears to be affirmative. A
fundamental example of entanglement in QFT is the case of the left
and right Rindler wedges, originally analyzed by Bisognano and Wichmann
\citep{Bisognano1975}. In fact, for an even-dimensional  massless and non-interacting
QFT, it has been shown \citep{Buchholz:1977ze,Olson:2010jy} that
entanglement exists between the operators within the future and past
light-cones, analogous to the entanglement between the left and right
Rindler wedges. This arises because, in such theories, every influence
propagates along only lightlike intervals, leading to both spacelike
commutativity and timelike commutativity. However, this is not generally
true in massive, interacting QFTs. Nevertheless, we can conclude that
entanglement across time does exist in certain special cases.

By the timelike tube theorem, the future and past light cones
are associated with distinct segments of an observer's worldline.
This provides a more physical interpretation of entanglement across
time: if an observer measures only massless, non-interacting quantum
fields, the observables measured in their future will be entangled
with those measured in their past. To explore entanglement across
time in the real world, a promising candidate is the pure Maxwell
field, which is closely related to Wheeler's delayed-choice experiment
with photons.
\begin{figure}[h]
\includegraphics[width=0.45\textwidth]{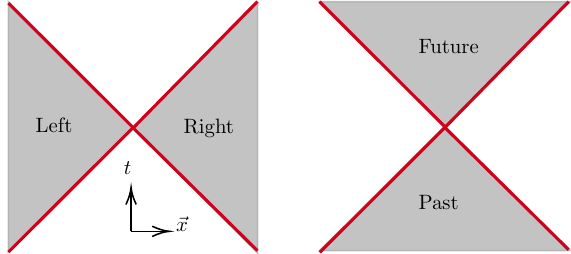}

\caption{The left/right Rindler wedges and the future/past light-cones are
illustrated, respectively.\label{fig:Rindler-Wedges}}
\end{figure}

\section*{Discussions}

In this letter, using known results in algebraic QFT, we proposed
a rigorous definition for timelike entanglement entropy in QFT of
general dimensions. The key insight is that timelike entanglement
entropy depends only on the timelike interval and is real-valued. 
However, our proposal may not apply to quantum field theories with defects or to generalized free field theories, where additivity or primitive causality is absent.

The algebra of observables has attracted significant attention recently
in quantum gravity and holography \citep{Leutheusser2023,Leutheusser2023a,Witten:2021unn,Penington:2023dql}.
In particular, the algebra of an observer's worldline plays a central
role in recent developments in quantum gravity \citep{Chandrasekaran:2022cip,Witten:2023qsv,Witten:2023xze}.
The timelike tube theorem has been generalized to QFTs in curved spacetime
\citep{Strohmaier2000,Strohmaier2023,Strohmaier2023a}, which, in
principle, allows us to define the timelike entanglement entropy in
the presence of gravity. 


\vspace*{3.0ex}
\begin{acknowledgments}
\paragraph*{Acknowledgments.} 
We deeply appreciate Edward Witten for pointing out that the strong subadditivity does not hold
on the timelike direction by the non-commutativity of the algebras. 
This fails the simple proof of $F-$theorem included in the first version of this paper.
This work is supported by NSFC (Grant No. 12275184).
\end{acknowledgments}

\bibliographystyle{unsrturl}
\bibliography{ref-2510-J}

\begin{thebibliography}{10}

\bibitem{Wang:2018jva}
Peng Wang, Houwen Wu, and Haitang Yang.
\newblock {Fix the dual geometries of $T\bar{T}$ deformed CFT$_{2}$ and highly
  excited states of CFT$_{2}$}.
\newblock {\em Eur. Phys. J. C}, 80(12):1117, 2020.
\newblock \href {http://arxiv.org/abs/1811.07758} {\path{arXiv:1811.07758}},
  \href {http://dx.doi.org/10.1140/epjc/s10052-020-08680-7}
  {\path{doi:10.1140/epjc/s10052-020-08680-7}}.

\bibitem{Liu:2022ugc}
Bowei Liu, Hao Chen, and Biao Lian.
\newblock {Entanglement entropy of free fermions in timelike slices}.
\newblock {\em Phys. Rev. B}, 110(14):144306, 2024.
\newblock \href {http://arxiv.org/abs/2210.03134} {\path{arXiv:2210.03134}},
  \href {http://dx.doi.org/10.1103/PhysRevB.110.144306}
  {\path{doi:10.1103/PhysRevB.110.144306}}.

\bibitem{Doi:2022iyj}
Kazuki Doi, Jonathan Harper, Ali Mollabashi, Tadashi Takayanagi, and Yusuke
  Taki.
\newblock {Pseudoentropy in dS/CFT and Timelike Entanglement Entropy}.
\newblock {\em Phys. Rev. Lett.}, 130(3):031601, 2023.
\newblock \href {http://arxiv.org/abs/2210.09457} {\path{arXiv:2210.09457}},
  \href {http://dx.doi.org/10.1103/PhysRevLett.130.031601}
  {\path{doi:10.1103/PhysRevLett.130.031601}}.

\bibitem{Doi:2023zaf}
Kazuki Doi, Jonathan Harper, Ali Mollabashi, Tadashi Takayanagi, and Yusuke
  Taki.
\newblock {Timelike entanglement entropy}.
\newblock {\em JHEP}, 05:052, 2 2023.
\newblock \href {http://arxiv.org/abs/2302.11695} {\path{arXiv:2302.11695}},
  \href {http://dx.doi.org/10.1007/JHEP05(2023)052}
  {\path{doi:10.1007/JHEP05(2023)052}}.

\bibitem{Milekhin:2025ycm}
Alexey Milekhin, Zofia Adamska, and John Preskill.
\newblock {Observable and computable entanglement in time}.
\newblock 2 2025.
\newblock \href {http://arxiv.org/abs/2502.12240} {\path{arXiv:2502.12240}}.

\bibitem{Witten:2023qsv}
Edward Witten.
\newblock {Algebras, regions, and observers.}
\newblock {\em Proc. Symp. Pure Math.}, 107:247--276, 2024.
\newblock \href {http://arxiv.org/abs/2303.02837} {\path{arXiv:2303.02837}},
  \href {http://dx.doi.org/10.1090/pspum/107/01954}
  {\path{doi:10.1090/pspum/107/01954}}.

\bibitem{Borchers1964}
Hans-J{\"u}rgen Borchers.
\newblock Field operators as $\mathbb{C}^{\infty}$ functions in spacelike
  directions.
\newblock {\em Il Nuovo Cimento (1955-1965)}, 33:1600--1613, 1964.

\bibitem{Haag1964}
Rudolf Haag and Daniel Kastler.
\newblock An algebraic approach to quantum field theory.
\newblock {\em Journal of Mathematical Physics}, 5(7):848--861, 1964.

\bibitem{Haag:1996hvx}
Rudolf Haag.
\newblock {\em Local Quantum Physics: Fields, Particles, Algebras}.
\newblock Theoretical and Mathematical Physics. Springer, Berlin, 1996.
\newblock \href {http://dx.doi.org/10.1007/978-3-642-61458-3}
  {\path{doi:10.1007/978-3-642-61458-3}}.

\bibitem{Casini:2020rgj}
Horacio Casini, Marina Huerta, Javier~M. Magan, and Diego Pontello.
\newblock {Entropic order parameters for the phases of QFT}.
\newblock {\em JHEP}, 04:277, 2021.
\newblock \href {http://arxiv.org/abs/2008.11748} {\path{arXiv:2008.11748}},
  \href {http://dx.doi.org/10.1007/JHEP04(2021)277}
  {\path{doi:10.1007/JHEP04(2021)277}}.

\bibitem{Casini:2021zgr}
Horacio Casini and Javier~M. Magan.
\newblock {On completeness and generalized symmetries in quantum field theory}.
\newblock {\em Mod. Phys. Lett. A}, 36(36):2130025, 2021.
\newblock \href {http://arxiv.org/abs/2110.11358} {\path{arXiv:2110.11358}},
  \href {http://dx.doi.org/10.1142/S0217732321300251}
  {\path{doi:10.1142/S0217732321300251}}.

\bibitem{Doplicher:1984zz}
Sergio Doplicher and Roberto Longo.
\newblock {Standard and split inclusions of von Neumann algebras}.
\newblock {\em Invent. Math.}, 75:493--536, 1984.
\newblock \href {http://dx.doi.org/10.1007/BF01388641}
  {\path{doi:10.1007/BF01388641}}.

\bibitem{Longo:2019pjj}
Roberto Longo and Feng Xu.
\newblock {Von Neumann Entropy in QFT}.
\newblock {\em Commun. Math. Phys.}, 381(3):1031--1054, 2021.
\newblock \href {http://arxiv.org/abs/1911.09390} {\path{arXiv:1911.09390}},
  \href {http://dx.doi.org/10.1007/s00220-020-03702-7}
  {\path{doi:10.1007/s00220-020-03702-7}}.

\bibitem{Narnhofer:2002Entanglement}
Heide Narnhofer.
\newblock Entanglement, split and nuclearity in quantum field theory.
\newblock {\em Rept. Math. Phys.}, 50(1):111--123, 2002.
\newblock \href {http://dx.doi.org/10.1016/S0034-4877(02)80048-9}
  {\path{doi:10.1016/S0034-4877(02)80048-9}}.

\bibitem{Casini:2009Reduced}
Horacio Casini and Marina Huerta.
\newblock Reduced density matrix and internal dynamics for multicomponent
  regions.
\newblock {\em Class. Quant. Grav.}, 26:185005, 2009.
\newblock \href {http://arxiv.org/abs/0903.5284} {\path{arXiv:0903.5284}},
  \href {http://dx.doi.org/10.1088/0264-9381/26/18/185005}
  {\path{doi:10.1088/0264-9381/26/18/185005}}.

\bibitem{Otani:2017Towards}
Yul Otani and Yoh Tanimoto.
\newblock Towards entanglement entropy with uv cutoff in conformal nets.
\newblock {\em Ann. Henri Poincare}, 19(6):1817--1842, 2018.
\newblock \href {http://arxiv.org/abs/1701.01186} {\path{arXiv:1701.01186}},
  \href {http://dx.doi.org/10.1007/s00023-018-0671-9}
  {\path{doi:10.1007/s00023-018-0671-9}}.

\bibitem{Hollands:2018Entanglement}
Stefan Hollands and Ko~Sanders.
\newblock {\em Entanglement Measures and Their Properties in Quantum Field
  Theory}, volume~34 of {\em SpringerBriefs in Mathematical Physics}.
\newblock Springer, Cham, 2018.
\newblock \href {http://arxiv.org/abs/1702.04924} {\path{arXiv:1702.04924}},
  \href {http://dx.doi.org/10.1007/978-3-319-94902-4}
  {\path{doi:10.1007/978-3-319-94902-4}}.

\bibitem{Witten:2018zxz}
Edward Witten.
\newblock {APS Medal for Exceptional Achievement in Research: Invited article
  on entanglement properties of quantum field theory}.
\newblock {\em Rev. Mod. Phys.}, 90(4):045003, 2018.
\newblock \href {http://arxiv.org/abs/1803.04993} {\path{arXiv:1803.04993}},
  \href {http://dx.doi.org/10.1103/RevModPhys.90.045003}
  {\path{doi:10.1103/RevModPhys.90.045003}}.

\bibitem{Dutta:2019gen}
Souvik Dutta and Thomas Faulkner.
\newblock {A canonical purification for the entanglement wedge cross-section}.
\newblock {\em JHEP}, 03:178, 2021.
\newblock \href {http://arxiv.org/abs/1905.00577} {\path{arXiv:1905.00577}},
  \href {http://dx.doi.org/10.1007/JHEP03(2021)178}
  {\path{doi:10.1007/JHEP03(2021)178}}.

\bibitem{Kudler-Flam:2023hkl}
Jonah Kudler-Flam, Samuel Leutheusser, Adel~A. Rahman, Gautam Satishchandran,
  and Antony~J. Speranza.
\newblock {Covariant regulator for entanglement entropy: Proofs of the
  Bekenstein bound and the quantum null energy condition}.
\newblock {\em Phys. Rev. D}, 111(10):105001, 2025.
\newblock \href {http://arxiv.org/abs/2312.07646} {\path{arXiv:2312.07646}},
  \href {http://dx.doi.org/10.1103/PhysRevD.111.105001}
  {\path{doi:10.1103/PhysRevD.111.105001}}.

\bibitem{Buchholz:1986Causal}
Detlev Buchholz and Eyvind~H. Wichmann.
\newblock Causal independence and the energy level density of states in local
  quantum field theory.
\newblock {\em Commun. Math. Phys.}, 106(2):321--344, 1986.
\newblock \href {http://dx.doi.org/10.1007/BF01454978}
  {\path{doi:10.1007/BF01454978}}.

\bibitem{Buchholz:1987Universal}
Detlev Buchholz, Claudio D'Antoni, and Klaus Fredenhagen.
\newblock The universal structure of local algebras.
\newblock {\em Commun. Math. Phys.}, 111(1):123--135, 1987.
\newblock \href {http://dx.doi.org/10.1007/BF01239019}
  {\path{doi:10.1007/BF01239019}}.

\bibitem{Buchholz:1990NuclearI}
Detlev Buchholz, Claudio D'Antoni, and Roberto Longo.
\newblock Nuclear maps and modular structures. i. general properties.
\newblock {\em J. Funct. Anal.}, 88(2):233--250, 1990.
\newblock \href {http://dx.doi.org/10.1016/0022-1236(90)90104-S}
  {\path{doi:10.1016/0022-1236(90)90104-S}}.

\bibitem{Buchholz:1990NuclearII}
Detlev Buchholz, Claudio D'Antoni, and Roberto Longo.
\newblock Nuclear maps and modular structures. ii. applications to quantum
  field theory.
\newblock {\em Commun. Math. Phys.}, 129(1):115--138, 1990.
\newblock \href {http://dx.doi.org/10.1007/BF02096782}
  {\path{doi:10.1007/BF02096782}}.

\bibitem{Buchholz:2007Nuclearity}
Detlev Buchholz, Claudio D'Antoni, and Roberto Longo.
\newblock Nuclearity and thermal states in conformal field theory.
\newblock {\em Commun. Math. Phys.}, 270(1):267--293, 2007.
\newblock \href {http://arxiv.org/abs/math-ph/0603083}
  {\path{arXiv:math-ph/0603083}}, \href
  {http://dx.doi.org/10.1007/s00220-006-0127-9}
  {\path{doi:10.1007/s00220-006-0127-9}}.

\bibitem{Morinelli:2018Conformal}
Vincenzo Morinelli, Yoh Tanimoto, and Mih{\'a}ly Weiner.
\newblock Conformal covariance and the split property.
\newblock {\em Commun. Math. Phys.}, 357(1):379--406, 2018.
\newblock \href {http://arxiv.org/abs/1609.02196} {\path{arXiv:1609.02196}},
  \href {http://dx.doi.org/10.1007/s00220-017-2961-3}
  {\path{doi:10.1007/s00220-017-2961-3}}.

\bibitem{Ceyhan:2025qrj}
Fikret Ceyhan and Thomas Faulkner.
\newblock {Bounds on CFT correlations from the thermal partition function}.
\newblock 10 2025.
\newblock \href {http://arxiv.org/abs/2510.24042} {\path{arXiv:2510.24042}}.

\bibitem{Verch:1993fs}
Rainer Verch.
\newblock {Nuclearity, split property and duality for the Klein-Gordon field in
  curved space-time}.
\newblock {\em Lett. Math. Phys.}, 29:297--310, 1993.
\newblock \href {http://dx.doi.org/10.1007/BF00750964}
  {\path{doi:10.1007/BF00750964}}.

\bibitem{DAntoni:2006Dirac}
Claudio D'Antoni and Stefan Hollands.
\newblock Nuclearity, local quasiequivalence and split property for dirac
  quantum fields in curved spacetime.
\newblock {\em Commun. Math. Phys.}, 261(1):133--159, 2006.
\newblock \href {http://arxiv.org/abs/math-ph/0106028}
  {\path{arXiv:math-ph/0106028}}, \href
  {http://dx.doi.org/10.1007/s00220-005-1398-2}
  {\path{doi:10.1007/s00220-005-1398-2}}.

\bibitem{Fewster:2015Split}
Christopher~J. Fewster.
\newblock The split property for locally covariant quantum field theories in
  curved spacetime.
\newblock {\em Lett. Math. Phys.}, 105(12):1633--1661, 2015.
\newblock \href {http://arxiv.org/abs/1501.02682} {\path{arXiv:1501.02682}},
  \href {http://dx.doi.org/10.1007/s11005-015-0798-2}
  {\path{doi:10.1007/s11005-015-0798-2}}.

\bibitem{Cao:2025els}
Xuchen Cao, Thomas Faulkner, and Zhencheng Wang.
\newblock {Gravitational Algebras with Two Areas}.
\newblock 12 2025.
\newblock \href {http://arxiv.org/abs/2512.04435} {\path{arXiv:2512.04435}}.

\bibitem{Borchers1961}
Hans-J{\"u}rgen Borchers.
\newblock {\"U}ber die vollst{\"a}ndigkeit lorentzinvarianter felder in einer
  zeitartigen r{\"o}hre.
\newblock {\em Il Nuovo Cimento (1955-1965)}, 19:787--793, 1961.

\bibitem{Araki1963}
Huzihiro Araki.
\newblock A generalization of borchers theorem.
\newblock {\em Helvetica Physica Acta (Switzerland)}, 36, 1963.

\bibitem{Haag:1962}
R.~Haag and B.~Schroer.
\newblock Postulates of quantum field theory.
\newblock {\em Journal of Mathematical Physics}, 3(2):248--256, 03 1962.
\newblock URL: \url{https://doi.org/10.1063/1.1703797}, \href
  {http://arxiv.org/abs/https://pubs.aip.org/aip/jmp/article-pdf/3/2/248/19054820/248_1_online.pdf}
  {\path{arXiv:https://pubs.aip.org/aip/jmp/article-pdf/3/2/248/19054820/248_1_online.pdf}},
  \href {http://dx.doi.org/10.1063/1.1703797} {\path{doi:10.1063/1.1703797}}.

\bibitem{Callan:1994py}
Curtis~G. Callan, Jr. and Frank Wilczek.
\newblock {On geometric entropy}.
\newblock {\em Phys. Lett. B}, 333:55--61, 1994.
\newblock \href {http://arxiv.org/abs/hep-th/9401072}
  {\path{arXiv:hep-th/9401072}}, \href
  {http://dx.doi.org/10.1016/0370-2693(94)91007-3}
  {\path{doi:10.1016/0370-2693(94)91007-3}}.

\bibitem{Ryu:2006bv}
Shinsei Ryu and Tadashi Takayanagi.
\newblock {Holographic derivation of entanglement entropy from AdS/CFT}.
\newblock {\em Phys. Rev. Lett.}, 96:181602, 2006.
\newblock \href {http://arxiv.org/abs/hep-th/0603001}
  {\path{arXiv:hep-th/0603001}}, \href
  {http://dx.doi.org/10.1103/PhysRevLett.96.181602}
  {\path{doi:10.1103/PhysRevLett.96.181602}}.

\bibitem{Ryu:2006ef}
Shinsei Ryu and Tadashi Takayanagi.
\newblock Aspects of holographic entanglement entropy.
\newblock {\em Journal of High Energy Physics}, 2006(08):045--045, aug 2006.
\newblock \href {http://dx.doi.org/10.1088/1126-6708/2006/08/045}
  {\path{doi:10.1088/1126-6708/2006/08/045}}.

\bibitem{Calabrese:2004eu}
Pasquale Calabrese and John~L. Cardy.
\newblock {Entanglement entropy and quantum field theory}.
\newblock {\em J. Stat. Mech.}, 0406:P06002, 2004.
\newblock \href {http://arxiv.org/abs/hep-th/0405152}
  {\path{arXiv:hep-th/0405152}}, \href
  {http://dx.doi.org/10.1088/1742-5468/2004/06/P06002}
  {\path{doi:10.1088/1742-5468/2004/06/P06002}}.

\bibitem{Calabrese:2009qy}
Pasquale Calabrese and John~L. Cardy.
\newblock {Entanglement entropy and conformal field theory}.
\newblock {\em J. Phys. A}, 42:504005, 2009.
\newblock \href {http://arxiv.org/abs/0905.4013} {\path{arXiv:0905.4013}},
  \href {http://dx.doi.org/10.1088/1751-8113/42/50/504005}
  {\path{doi:10.1088/1751-8113/42/50/504005}}.

\bibitem{Cardy1986}
John~L Cardy.
\newblock Operator content of two-dimensional conformally invariant theories.
\newblock {\em Nuclear Physics B}, 270:186--204, 1986.

\bibitem{Cardy:2004hm}
John~L. Cardy.
\newblock {Boundary conformal field theory}.
\newblock 11 2004.
\newblock \href {http://arxiv.org/abs/hep-th/0411189}
  {\path{arXiv:hep-th/0411189}}.

\bibitem{Cardy:2016fqc}
John Cardy and Erik Tonni.
\newblock {Entanglement hamiltonians in two-dimensional conformal field
  theory}.
\newblock {\em J. Stat. Mech.}, 1612(12):123103, 2016.
\newblock \href {http://arxiv.org/abs/1608.01283} {\path{arXiv:1608.01283}},
  \href {http://dx.doi.org/10.1088/1742-5468/2016/12/123103}
  {\path{doi:10.1088/1742-5468/2016/12/123103}}.

\bibitem{Brown1986}
J.~David Brown and M.~Henneaux.
\newblock Central charges in the canonical realization of asymptotic
  symmetries: An example from three-dimensional gravity.
\newblock {\em Commun. Math. Phys.}, 104:207--226, 1986.
\newblock \href {http://dx.doi.org/10.1007/BF01211590}
  {\path{doi:10.1007/BF01211590}}.

\bibitem{Jiang:2024ijx}
Xin Jiang, Peng Wang, Houwen Wu, and Haitang Yang.
\newblock {Alternative to purification in conformal field theory}.
\newblock {\em Phys. Rev. D}, 111(2):L021902, 2025.
\newblock \href {http://arxiv.org/abs/2406.09033} {\path{arXiv:2406.09033}},
  \href {http://dx.doi.org/10.1103/PhysRevD.111.L021902}
  {\path{doi:10.1103/PhysRevD.111.L021902}}.

\bibitem{Jiang:2024hjz}
Xin Jiang, Peng Wang, Houwen Wu, and Haitang Yang.
\newblock {How Einstein{\textquoteright}s equations emerge from CFT2}.
\newblock {\em Phys. Rev. D}, 112(8):8, 2025.
\newblock \href {http://arxiv.org/abs/2410.19711} {\path{arXiv:2410.19711}},
  \href {http://dx.doi.org/10.1103/zg5x-34mn} {\path{doi:10.1103/zg5x-34mn}}.

\bibitem{Jiang:2025tqu}
Xin Jiang, Peng Wang, Houwen Wu, and Haitang Yang.
\newblock {Mixed state entanglement entropy in CFT}.
\newblock {\em JHEP}, 09:133, 2025.
\newblock \href {http://arxiv.org/abs/2501.08198} {\path{arXiv:2501.08198}},
  \href {http://dx.doi.org/10.1007/JHEP09(2025)133}
  {\path{doi:10.1007/JHEP09(2025)133}}.

\bibitem{Jiang:2025jnk}
Xin Jiang and Haitang Yang.
\newblock {Entanglement entropy of conformal field theory in all dimensions}.
\newblock {\em JHEP}, 01:015, 2026.
\newblock \href {http://arxiv.org/abs/2506.02786} {\path{arXiv:2506.02786}},
  \href {http://dx.doi.org/10.1007/JHEP01(2026)015}
  {\path{doi:10.1007/JHEP01(2026)015}}.

\bibitem{Calabrese:2009ez}
Pasquale Calabrese, John Cardy, and Erik Tonni.
\newblock {Entanglement entropy of two disjoint intervals in conformal field
  theory}.
\newblock {\em J. Stat. Mech.}, 0911:P11001, 2009.
\newblock \href {http://arxiv.org/abs/0905.2069} {\path{arXiv:0905.2069}},
  \href {http://dx.doi.org/10.1088/1742-5468/2009/11/P11001}
  {\path{doi:10.1088/1742-5468/2009/11/P11001}}.

\bibitem{Calabrese:2010he}
Pasquale Calabrese, John Cardy, and Erik Tonni.
\newblock {Entanglement entropy of two disjoint intervals in conformal field
  theory II}.
\newblock {\em J. Stat. Mech.}, 1101:P01021, 2011.
\newblock \href {http://arxiv.org/abs/1011.5482} {\path{arXiv:1011.5482}},
  \href {http://dx.doi.org/10.1088/1742-5468/2011/01/P01021}
  {\path{doi:10.1088/1742-5468/2011/01/P01021}}.

\bibitem{Bisognano1975}
Joseph~J Bisognano and Eyvind~H Wichmann.
\newblock On the duality condition for a hermitian scalar field.
\newblock {\em Journal of Mathematical Physics}, 16(4):985--1007, 1975.
\newblock \href {http://dx.doi.org/10.1063/1.522605}
  {\path{doi:10.1063/1.522605}}.

\bibitem{Buchholz:1977ze}
D.~Buchholz.
\newblock {On the Structure of Local Quantum Fields with Nontrivial
  Interaction}.
\newblock pages 146--153, 1977.

\bibitem{Olson:2010jy}
S.~Jay Olson and Timothy~C. Ralph.
\newblock {Entanglement between the future and past in the quantum vacuum}.
\newblock {\em Phys. Rev. Lett.}, 106:110404, 2011.
\newblock \href {http://arxiv.org/abs/1003.0720} {\path{arXiv:1003.0720}},
  \href {http://dx.doi.org/10.1103/PhysRevLett.106.110404}
  {\path{doi:10.1103/PhysRevLett.106.110404}}.

\bibitem{Leutheusser2023}
Samuel Leutheusser and Hong Liu.
\newblock {Causal connectability between quantum systems and the black hole
  interior in holographic duality}.
\newblock {\em Phys. Rev. D}, 108(8):086019, 2023.
\newblock \href {http://arxiv.org/abs/2110.05497} {\path{arXiv:2110.05497}},
  \href {http://dx.doi.org/10.1103/PhysRevD.108.086019}
  {\path{doi:10.1103/PhysRevD.108.086019}}.

\bibitem{Leutheusser2023a}
Samuel Leutheusser and Hong Liu.
\newblock {Emergent times in holographic duality}.
\newblock {\em Phys. Rev. D}, 108(8):086020, 2023.
\newblock \href {http://arxiv.org/abs/2112.12156} {\path{arXiv:2112.12156}},
  \href {http://dx.doi.org/10.1103/PhysRevD.108.086020}
  {\path{doi:10.1103/PhysRevD.108.086020}}.

\bibitem{Witten:2021unn}
Edward Witten.
\newblock {Gravity and the crossed product}.
\newblock {\em JHEP}, 10:008, 2022.
\newblock \href {http://arxiv.org/abs/2112.12828} {\path{arXiv:2112.12828}},
  \href {http://dx.doi.org/10.1007/JHEP10(2022)008}
  {\path{doi:10.1007/JHEP10(2022)008}}.

\bibitem{Penington:2023dql}
Geoff Penington and Edward Witten.
\newblock {Algebras and States in JT Gravity}.
\newblock 1 2023.
\newblock \href {http://arxiv.org/abs/2301.07257} {\path{arXiv:2301.07257}}.

\bibitem{Chandrasekaran:2022cip}
Venkatesa Chandrasekaran, Roberto Longo, Geoff Penington, and Edward Witten.
\newblock {An algebra of observables for de Sitter space}.
\newblock {\em JHEP}, 02:082, 2023.
\newblock \href {http://arxiv.org/abs/2206.10780} {\path{arXiv:2206.10780}},
  \href {http://dx.doi.org/10.1007/JHEP02(2023)082}
  {\path{doi:10.1007/JHEP02(2023)082}}.

\bibitem{Witten:2023xze}
Edward Witten.
\newblock {A background-independent algebra in quantum gravity}.
\newblock {\em JHEP}, 03:077, 2024.
\newblock \href {http://arxiv.org/abs/2308.03663} {\path{arXiv:2308.03663}},
  \href {http://dx.doi.org/10.1007/JHEP03(2024)077}
  {\path{doi:10.1007/JHEP03(2024)077}}.

\bibitem{Strohmaier2000}
Alexander {Strohmaier}.
\newblock {On the local structure of the Klein-Gordon field on curved
  spacetimes}.
\newblock {\em arXiv e-prints}, pages math--ph/0008043, August 2000.
\newblock \href {http://arxiv.org/abs/math-ph/0008043}
  {\path{arXiv:math-ph/0008043}}, \href
  {http://dx.doi.org/10.48550/arXiv.math-ph/0008043}
  {\path{doi:10.48550/arXiv.math-ph/0008043}}.

\bibitem{Strohmaier2023}
Alexander {Strohmaier} and Edward {Witten}.
\newblock {Analytic states in quantum field theory on curved spacetimes}.
\newblock {\em arXiv e-prints}, page arXiv:2302.02709, February 2023.
\newblock \href {http://arxiv.org/abs/2302.02709} {\path{arXiv:2302.02709}},
  \href {http://dx.doi.org/10.48550/arXiv.2302.02709}
  {\path{doi:10.48550/arXiv.2302.02709}}.

\bibitem{Strohmaier2023a}
Alexander {Strohmaier} and Edward {Witten}.
\newblock {The Timelike Tube Theorem in Curved Spacetime}.
\newblock {\em arXiv e-prints}, page arXiv:2303.16380, March 2023.
\newblock \href {http://arxiv.org/abs/2303.16380} {\path{arXiv:2303.16380}},
  \href {http://dx.doi.org/10.48550/arXiv.2303.16380}
  {\path{doi:10.48550/arXiv.2303.16380}}.

\end{thebibliography}

\end{document}